\documentclass[aps,twocolumn,showpacs]{revtex4}
\usepackage{graphicx,color}
\newcommand{\ds}{ _{\downarrow}}
\newcommand{\us}{ _{\uparrow}}
\newcommand{\up}{\uparrow}
\newcommand{\down}{\downarrow}
\newcommand{\Li}{\mathop{\textrm{Li}}}

\begin{document}
\draft \title{Highly polarized Fermi gases:  One-dimensional case}
\author{S. Giraud and R. Combescot} \address{Laboratoire de
Physique Statistique, Ecole Normale Sup\'erieure, 24 rue Lhomond,
75231 Paris Cedex 05, France}
\date{Received \today}

\begin{abstract}
We consider the problem of a single particle interacting with $N$ identical fermions, at zero temperature and in one dimension.
We calculate the binding energy as well as the effective mass of the single particle.
We use an approximate method developed in the three-dimensional case, where
the Hilbert space for the excited states of the $N$ fermions is restricted to have at most two particle-hole pairs. 
When the mass of the single particle is equal to the fermion mass, we find an excellent agreement
with the exact results of McGuire. When the mass of the single particle is infinite, we solve exactly the problem
and find again excellent agreement between approximate results and exact ones.
This overall agreement in one dimension gives a strong validation
for the approximate method applied in three dimensions. Moreover it shows that our approximate treatment is excellent 
for the one dimensional problem in the general case with respect to the mass of the single particle.
\end{abstract}

\maketitle

\section{introduction}
The field of ultracold fermionic atoms has seen during the last ten years an impressive amount of experimental and theoretical activity \cite{gps}. These systems provide experimental realization of various many-body systems that for a long time have been considered as purely theoretical models \cite{gaudin,mcg}. For instance, the 1D Fermi gas can be experimentally realized by tightly confining the atomic cloud in the radial directions and weakly confining it along the axial direction. Its behavior can be characterized to a very good approximation by an effective 1D coupling constant $g$ \cite{ols}. One major experimental achievement of the last few years was the realization of the BEC-BCS crossover by driving a mixture of two fermionic populations with an attractive interaction through a Feshbach resonance. On the BEC side of the crossover, a bound state appears between particles of different species, while on the BCS side, pairs exist only in the presence of the whole Fermi sea. In one dimension, the situation is quite different, because any attractive interaction produces a bound state. However, an analog of the 3D crossover can be achieved in a quasi-1D trap where one can vary the effective coupling constant $g$ from $-\infty$ to $+\infty$ by detuning the 3D scattering length \cite{afrzt}.

Recently attention has focused on imbalanced systems and in particular on the strongly imbalanced regime. The extreme situation corresponds to the case of a single $\downarrow$-spin particle interacting with a $\uparrow$-spin Fermi sea. The effective quasiparticle parameters, namely, the binding energy $E_b$ of the $\downarrow$-spin atom and its effective mass $m^\ast$, are known to be of major interest for the analysis of the phase diagram \cite{lrgs,pg}. In the three-dimensional case, analytical and MC calculations have been done \cite{lrgs,fred,crlc,pg,ps1,ps2}. 
Quite surprisingly a simple variational calculation \cite{fred,crlc} gives results remarkably close to MC calculations.
Very recently  \cite{cg} we have explained this puzzling feature and shown that this simple calculation
is just the first step in a series of approximations converging extremely rapidly toward the exact result.
These approximations correspond to restrict the Hilbert space to states with increasing number of
particle-hole excitations. The very fast convergence is due to almost perfect destructive interferences occurring
when more than one particle-hole excitation is present. As a result  \cite{cg} a variational calculation
taking into account at most two particle-hole excitations provides already an essentially exact solution to the problem. In this paper, we use the same approximation to address the 1D problem, having mostly in mind the case of attractive interaction. However we consider also
the case of repulsive interaction. In the case of equal masses, we can compare our results with the exact solution obtained by McGuire \cite{mcg}. The excellent agreement we find provides a further check of the reliability of our approach.

\section{general calculation}

We consider the problem of a single $\downarrow$-spin particle with mass $m\ds$ in the presence of a 1D Fermi sea, with Fermi momentum $k_F$ and density $n\us=k_F/\pi$, of $\uparrow$-spins with mass $m\us$. In the limit of short-range interactions, we can write the Hamiltonian of the system as:
\begin{eqnarray}\label{ham}
H&=&H_c+V \\
H_c&=&\sum_{Q}E(Q)b^{\dag}_{Q}b_{Q}+ \sum_{k}\epsilon _{k}c^{\dag}_{k}c_{k}\\
V&=&g \sum_{K K' Q Q'}\delta_{K K' Q Q'}c^{\dag}_{K}c_{K'}b^{\dag}_{Q}b_{Q'}
\end{eqnarray}
where $\epsilon_{k} =k^{2}/2m\us$, $E(Q)=Q^2/2m\ds$, and $c_{k}$ and $c^{\dag}_{k}$ are annihilation and creation operators for $\uparrow$-spin atoms while $b_{k}$ and $b^{\dag}_{k}$ are for the $\downarrow$-spin atom. The Kronecker symbol $\delta_{K K' Q Q'}$ ensures momentum conservation in the scattering. Contrary to the three-dimensional case \cite{cg}, the representation of the interaction potential by a Dirac distribution is mathematically well-defined, since there are no ultraviolet divergences. The coupling constant $g$ is either negative or positive, and there is no need to let it go to zero. It is related to the value of the one-dimensional scattering length by $g=-1/(m_ra)$, where $m_r=m\ds m\us/(m\ds +m\us)$ is the reduced mass (we have set $\hbar=1$).

Following the same idea as in Ref. \cite{cg}, we write, for a system of total momentum $p$, the many-body wave function as: 
\begin{eqnarray}\label{trial}
|\psi\rangle&=&\alpha _0 b^{\dag}_p|0\rangle+ \sum_{kq}\alpha _{kq}b^{\dag}_{p+q-k}c^{\dag}_{k}c_{q}|0\rangle \\\nonumber
&&+\frac{1}{4}\sum_{kk'qq'}\alpha _{kk'qq'}b^{\dag}_{p+q+q'-k-k'}c^{\dag}_{k}c^{\dag}_{k'}c_{q}c_{q'}|0\rangle
\end{eqnarray}
where $|0\rangle=\prod_{|q|<k_F} c^{\dag}_{q}\,|vac\rangle$ is the noninteracting Fermi sea of $\uparrow$-spins and the sums on $q$ and $k$ are implicitly limited to $|q|<k_F$ and $|k|>k_F$. This state is a superposition of states with up to two particle-hole pairs. In the first term, the $\up$-spin free Fermi sea is in its ground state and the $\downarrow$-spin atom carries the momentum $p$, while the two other terms correspond to excited states with creation of one or two particle-hole pairs in the Fermi sea, the $\down$-spin atom carrying the rest of the momentum. In the third term, the coefficients $\alpha _{kk'qq'}$ are antisymmetric with respect to the exchange of their arguments ($\alpha_{kk'qq'}=-\alpha_{k'kqq'}=-\alpha_{kk'q'q}=\alpha_{k'kq'q}$) and the factor $1/4$ corrects multiple counting.

Now, we write the Schr\"odinger equation $H|\psi\rangle=E|\psi\rangle$ and project it on the subspace corresponding to Eq.(\ref{trial}). This yields a set of three coupled equations. The two first equations are obtained by projecting onto the full Fermi sea and the Fermi sea with a single particle-hole pair:
\begin{eqnarray}\label{eq0}
&&\hspace{-7mm}-g^{-1}E^{(0)}_{p}\alpha _0\!=\!\sum_{kq}\alpha _{kq}\\
\label{eq1}
&&\hspace{-7mm}-g^{-1}E^{(1)}_{p,kq}\alpha _{kq}\!=\!\alpha _0\!+\!\! \sum_{K} \alpha _{Kq}-
\!\!\sum_{Q} \alpha _{kQ}-\!\! \sum_{KQ}\alpha _{kKqQ}
\end{eqnarray}
where we have introduced $E^{(0)}_{p}=-E+gn\us+E(p)$, and $E^{(1)}_{p,kq}=-E+gn\us+E(p+q-k)+\epsilon_k-\epsilon_q$. Here $E$ denotes  the energy measured from the energy of the free Fermi sea.  Note that, if we omit the last term of Eq.(\ref{eq1}), these two first equations would be decoupled from the third one and this would correspond to the variational calculation \cite{crlc} with a single particle-hole pair. The next equation is:
\begin{eqnarray}
&&\hspace{-3mm}-g^{-1}E^{(2)}_{p,kk'qq'}\alpha _{kk'qq'}\!=\!-\alpha _{kq}\!-\alpha _{k'q'}\!+\alpha _{kq'}\!+\alpha _{k'q} \label{eq2} \\
&&\hspace{3mm} + \!\sum_{K} \alpha _{Kk'qq'}\!+\!\sum_{K} \alpha _{kKqq'}\!-\!\!\sum_{Q} \alpha _{kk'Qq'}\!-\!\!\sum_{Q} \alpha _{kk'qQ} \nonumber
\end{eqnarray}
where $E^{(2)}_{p,kk'qq'}=-E+gn\us+E(p+q+q'-k-k')+\epsilon_k+\epsilon_{k'}-\epsilon_q-\epsilon_{q'}$. Here, since we have restricted the Hilbert space to contain at most two particle-hole pairs, we have omitted the overlap with states containing three particle-hole pairs. This restriction corresponds to perform a variational calculation. For $p=0$, we have $E=-E_b$, while the variation of $E$ for small $p$ gives the effective mass. We consider first the calculation of the binding energy.

It is first worth noticing that, for attractive interaction, the two limiting cases, namely $g\to 0_{-}$ and $g\to -\infty$, are already well described by the first order approximation,
corresponding to retain only Eq.(\ref{eq0}) and Eq.(\ref{eq1}), with the last term in the right-hand side of Eq.(\ref{eq1}) omitted. 
In the weak coupling limit
$g\to 0_{-}$, Eq.(\ref{eq0}) gives to lowest order the expected result, that is the mean-field interaction energy $E=gn\us$. Then Eq.(\ref{eq1}) provides the correction expected from second order perturbation theory, namely
$E=gn\us-m\us g^2[1/8+(\Li_2(\alpha)-\Li_2(-\alpha))/(2\pi^2)]$, where $\alpha=(m\ds-m\us)/(m\ds+m\us)$ and $\Li_2(z)=\sum_{n\geq 1}z^n/n^2$ is the dilogarithm function. In the case of equal masses, the result is identical to the one obtained by McGuire \cite{mcg} in the same approximation. 
 
In the strong coupling regime, we can check that the binding energy of the molecular state $-E=E_b=m_rg^2/2$ is indeed solution to dominant order. In this regime, since $q$ is limited to $|q|<k_F$, the dependence on $q$ is negligible in $E^{(1)}_{0,kq}$. Now we can sum Eq.(\ref{eq1}) over $q$. This gives $-g^{-1}\left[E_b+k^2/(2m_r)\right]\sum_q\alpha_{kq}=\alpha_0k_F/\pi+\sum_{Kq}\alpha_{Kq}$.
Then we can divide this equation by $\left[E_b+k^2/(2m_r)\right]$ and sum over $k$. To dominant order we find $\sum_{k}\left[E_b+k^2/(2m_r)\right]^{-1}=\sqrt{m_r/(2E_b)}$, leading to $\left[1+g^{-1}\sqrt{2E_b/m_r}\right]\sum_{kq}\alpha _{kq}=-\alpha_0k_F/\pi$. Inserting this relation in Eq.(\ref{eq0}) gives $1+g^{-1}\sqrt{2E_b/m_r}= gk_F/(\pi E_b)$ to dominant order. This is indeed consistent with $E_b \approx m_rg^2/2$.

In the general case, these equations can be numerically solved by successive iterations to provide the binding energy $E_b$ 
and the effective mass $m^{\ast}$. 
This is done in the same way as for the three dimensional treatment, but obviously the numerical work is much lighter since one needs to perform
only one-dimensional integrals. On the other hand, in contrast with the 3D situation, the third term in the right-hand side of Eq.(\ref{eq1}) and the
last two terms in the right-hand side of Eq.(\ref{eq2}) do not disappear. However since we use an iterative method this does not make a major
problem. 

\section{Equal masses}
We first investigate the situation of equal masses $m\us=m\ds=m$. In this case, the exact solution of this problem has been obtained by McGuire \cite{mcg} using the Bethe ansatz (with a small sign mistake in the last two equations of page 126 and in the expression for $m^\ast$). With our notations (we have the opposite sign convention for the interaction) the correct expressions are:
\begin{eqnarray}\label{muex}
\frac{E_b}{E_F}&=&-\frac{2}{\pi}\left[y-\frac{\pi}{2}y^2+(1+y^2)\arctan y\right]\\
\frac{m^\ast}{m}&=&\frac{\left(1-\frac{2}{\pi}\arctan y\right)^2}{1-\frac{2}{\pi}\left[\arctan y+\frac{y}{1+y^2}\right]}\label{mex}
\end{eqnarray}
where $y=mg/2k_F$ and $E_F=k_F^2/2m$ is the Fermi energy of the $\up$ atoms. They allow us to check the quality of the
convergence of our successive approximations to the exact many-body solution.

\begin{figure}
\includegraphics[width=\linewidth]{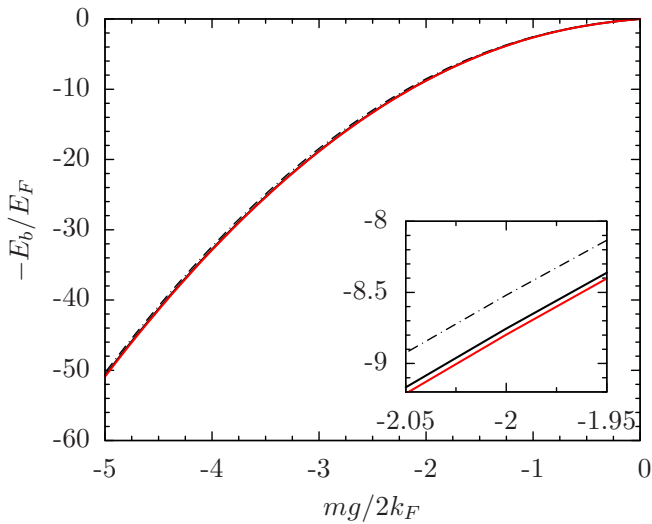}
\caption{\label{fig1} (Color online) Reduced $\down$-atom binding energy $E_b$ as a function of $mg/2k_F$ for equal masses $m\us=m\ds=m$ and attractive interactions $g<0$. From top to bottom: numerical results for the first (dashed-dotted line) and second (solid thick line) order approximations and exact  solution Eq.(\ref{muex}) (solid red line). The inset shows a zoom of the region around $mg/2k_F=-2$ (we have chosen a value larger than the one discussed in the text to make it easier to see the difference on the figure). }
\end{figure}

We begin by considering the case of attractive interactions $g<0$, which is the more interesting one in relation with the 3D case \cite{cg}. The overall result of the comparison is displayed on Fig.\ref{fig1}. As it was indicated in Ref.\cite{crlc} and is clearly seen in Fig.\ref{fig1}, the lowest order approximation for the binding energy is already in good agreement with the exact solution and the two curves are barely distinguishable on the graph. However, the second order approximation improves it significantly, leading to results completely undistinguishable from the exact result on the graph. To be more specific, let us consider the case for $mg/k_F=-1$. The exact result is $E_b=0.9373\,E_F$. We may consider the Hartreee
approximation $E_b=-gn\us=-(2/\pi )(mg/k_F)\,E_F$ as a zeroth order approximation. In the present case it gives $E_b=0.6366\,E_F$, which means a 
$3.2\; 10^{-1}$ relative error compared to the exact result. 
The first  order approximation gives $E_b=0.9242\,E_F$, that is a $1.4\; 10^{-2}$ relative error. Finally we find for our second order
approximation $E_b=0.9361\,E_F$, which corresponds to only a $1.3 \;10^{-3}$ relative error. 
We see that the convergence is very fast, which gives very strong support to
the idea that the result obtained in the three-dimensional case is nearly the exact solution. Actually this is even more so because the
convergence in 1D is expected to be slower than in 3D. Indeed the rapidity of this convergence is related \cite{cg} to the effective phase space
corresponding to particle wavevectors $k > k_F$, compared to the phase space corresponding to the hole wavevectors $q < k_F$. The ratio
between the volumes of these phase spaces is clearly an increasing function of space dimensionality.

\begin{figure}
\includegraphics[width=\linewidth]{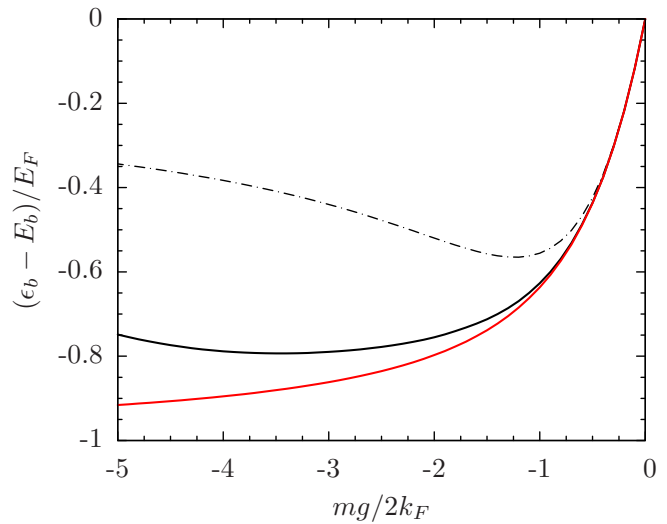}
\caption{\label{fig1b} (Color online) Reduced $\down$-atom binding energy $E_b$ as a function of $mg/2k_F$ for equal masses $m\us=m\ds=m$. Same conventions as in Fig.\ref{fig1} are used.}
\end{figure}

Naturally, for large values of the coupling constant, all the results look very close on Fig.\ref{fig1} because they are near
the large asymptotic value $-\epsilon _b/E_F=-\,2\,(mg/2k_F)^2$. In order to make the differences easier to see, we have plotted
in Fig.\ref{fig1b} the difference between the results and this asymptotic expression.

\begin{figure}
\includegraphics[width=\linewidth]{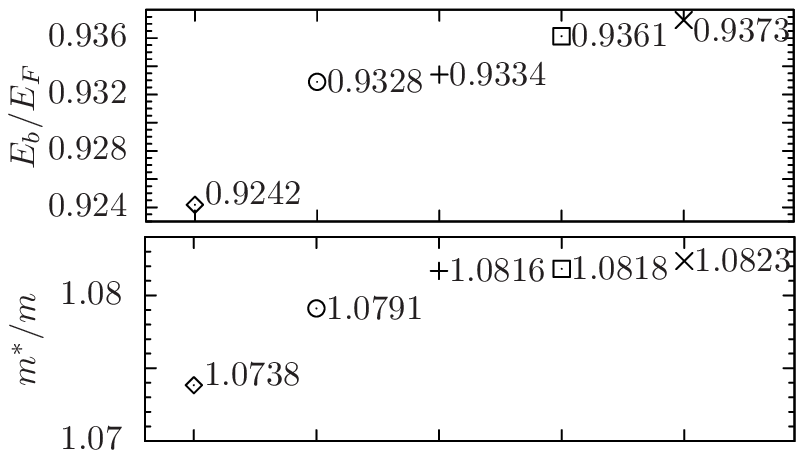}
\caption{\label{fig2} Reduced $\down$-atom binding energy $E_b$ and effective mass $m^\ast$, for equal masses $m\us=m\ds=m$ and $mg/k_F=-1$. Successive approximations to the exact many-body problem discussed in the text are applied. Diamond: first order approximation. Circle: second order with $qq'=00$. Plus: second order with $qq'=q0$. Square: second order with no $q$ approximation. Cross: exact solution.}
\end{figure}

We may also apply the same cascade of successive approximations as Ref.\cite{cg}. This leads to the approximations 
$E^{(1)}_{p,kq}\simeq E^{(1)}_{p,k0}$ in the first order equations and $E^{(2)}_{p,kk'qq'}\simeq E^{(2)}_{p,kk'00}$ or $E^{(2)}_{p,kk'qq'}\simeq E^{(2)}_{p,kk'q0}$ in the second order equations. The results are displayed in Fig.\ref{fig2} for $mg/k_F=-1$.

\begin{figure}
\includegraphics[width=\linewidth]{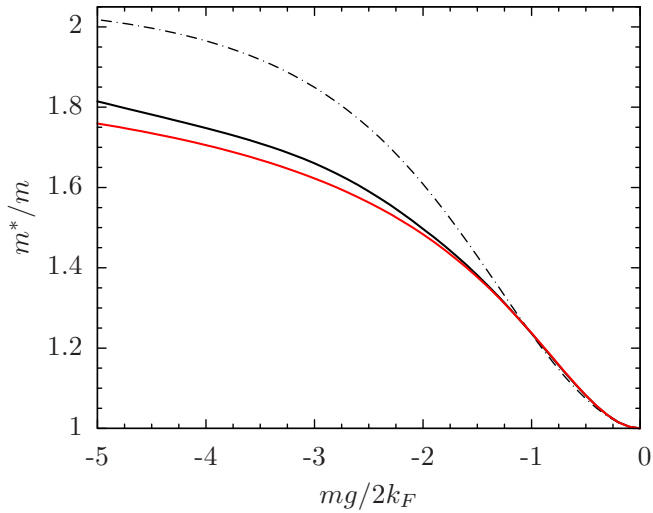}
\caption{\label{fig3} (Color online) Relative effective mass $m^\ast/m$ as a function of $mg/2k_F$ for equal masses $m\us=m\ds=m$. Same conventions as in Fig.\ref{fig1} are used.}
\end{figure}

Finally, the relative effective mass $m^\ast/m$ is plotted in Fig.\ref{fig3}. Naturally, in the weak coupling limit, $m^\ast$ approaches $m$ since the quasiparticle has the effective mass of the free $\downarrow$-spin atom. In the strong coupling regime, $m^\ast$ approaches the mass $2m$ of the two bound particles forming a dimer, as it is expected physically. The main quantitative feature of our results is that, contrary to the binding energy, 
the effective mass is more sensitive to our approximations as one could expect, especially in the strong coupling regime. 
The first order approximation is substantially improved by our inclusion of two particle-hole excitations. As above, this second order approximation is fairly close to the exact solution. We note that although our variational calculation gives naturally an upper bound for the energy $E$, the obtained effective mass is not a lower bound of the exact one since it is given by the variation of $E$.

\begin{figure}
\includegraphics[width=\linewidth]{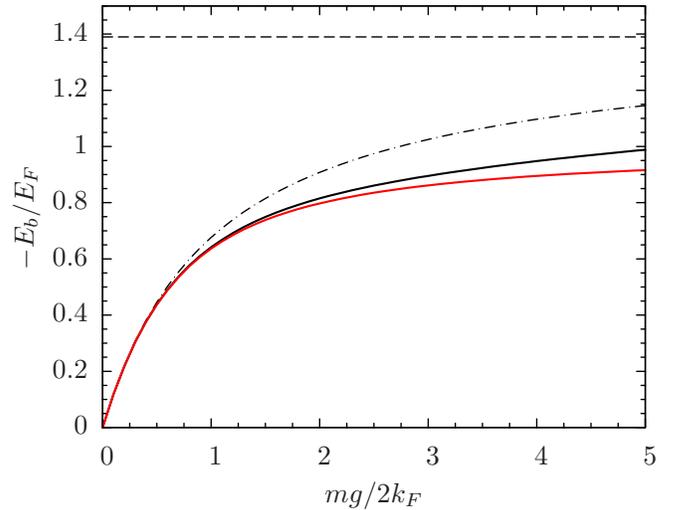}
\caption{\label{fig4} (Color online) Reduced $\down$-atom binding energy $E_b$ as a function of $mg/2k_F$ for equal masses $m\us=m\ds=m$ and repulsive interactions $g>0$. Same conventions as in Fig.\ref{fig1} are used. In the strong repulsive limit $g\to\infty$, first and second order approximations give $E=-E_b=1.389\,E_F$ (indicated by the dashed line) while the exact result is $E=-E_b=E_F$.}
\end{figure}

\begin{figure}
\includegraphics[width=\linewidth]{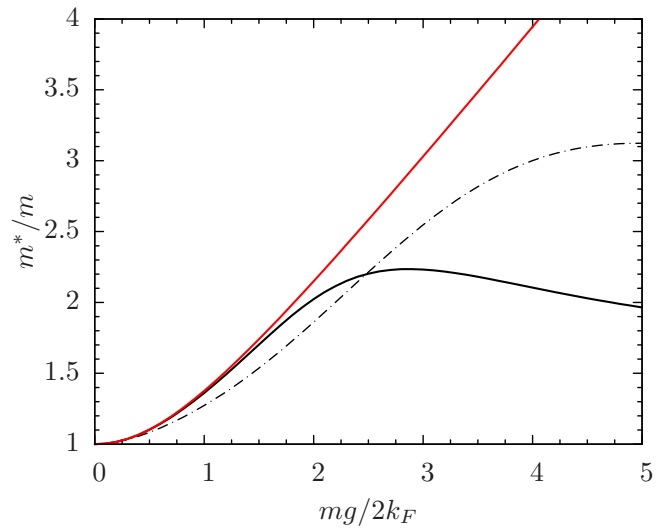}
\caption{\label{fig5} (Color online) Relative effective mass $m^\ast/m$ as a function of $mg/2k_F$ for equal masses $m\us=m\ds=m$ and repulsive interactions $g>0$. Same conventions as in Fig.\ref{fig1}.}
\end{figure}

It is also of interest to investigate the case of repulsive interactions $g>0$, since it is no more difficult to calculate than the attractive one. The result for the energy is displayed in Fig.\ref{fig4}. One sees that for small couplings the first and second order approximations are still excellent, which can merely be understood by continuity from the attractive case with small coupling, or the fact that in this domain standard perturbation theory is expected to be good. However for larger coupling the first order approximation becomes markedly less satisfactory. The
second order approximation remains quite good, but also with a trend toward a deterioration of the agreement. This feature is actually easy to understand. For strongly attractive interaction, the large negative energy $E$ makes $E^{(1)}_{p,kq}$ and $E^{(2)}_{p,kk'qq'}$ large and positive. As a result they depend weakly on the $q$ (hole) variables, which is the basic reason \cite{cg} for the success of the approximations. On the other hand for repulsive interactions, $E$ becomes positive and $E^{(1)}_{p,kq}$ and $E^{(2)}_{p,kk'qq'}$ are smaller, and accordingly their relative change under variation of the $q$ variables becomes more important. As expected under these circumstances, the agreement for the effective mass, shown in Fig.\ref{fig5}, is very rapidly unsatisfactory for the first order approximation. The second order approximation stays quite good until $mg/2k_F \sim 1.5$, but the disagreement increases quite markedly when the coupling is increased much further. These results are interesting since they come as a confirmation of the basic reason \cite{cg} for the quality of our approximation in the attractive domain.

In the strong repulsive limit $g\to\infty$, the binding energies given by the first and second order approximations are seen to converge to the same value. This limiting
value can be determined analytically by performing a $1/g$ expansion in the following way 
(we make the calculation directly for a general value of the mass ratio $r=m\ds/m\us$ since
this does not make any problem). We consider the equations of the first order approximation, that is 
Eq.(\ref{eq0}) and Eq.(\ref{eq1}) with the last term in the right-hand side omitted. We set for simplicity $\alpha_0=1$ and expand $\alpha_{kq}$ in powers of $1/y$ 
where $y=m_rg/k_F$: 
\begin{eqnarray}\label{}
\alpha_{kq}=\alpha_{kq}^{(0)}+\alpha_{kq}^{(1)}/y+\cdots
\end{eqnarray}
To lowest order, Eq.(\ref{eq0}) and Eq.(\ref{eq1}) give:
\begin{eqnarray}\label{eq11}
\hspace{-7mm}-n\us&=&\!\sum_{kq}\alpha _{kq}^{(0)}\\
-n\us\alpha_{kq}^{(0)}&=&1+\sum_{K} \alpha _{Kq}^{(0)}-\sum_{Q}\alpha _{kQ}^{(0)}
\label{eq12}
\end{eqnarray}
Since in Eq.(\ref{eq12}) the sum on $K$ in the right-hand side converges, this implies that we have $\alpha_{kq}^{(0)}\to 0$ for $k\to\infty$. Taking the limit $k\to\infty$ in this equation, we obtain $\sum_{K} \alpha _{Kq}^{(0)}=-1$ and accordingly, from Eq.(\ref{eq12}), $\alpha_{kq}^{(0)} \equiv \alpha_{k}^{(0)}$ does not depend on $q$. Note that, after summation over $q$, this finding is completely consistent with Eq.(\ref{eq11}). Now to first order in $1/y$, we find:
\begin{eqnarray}\label{eq13}
E_b=-\frac{k_F}{m_r}\,\sum_{kq}\alpha _{kq}^{(1)}\hspace{27mm}\\
\hspace{-7mm}(E_b+E(q-k)+\epsilon_k-\epsilon_q)\alpha _{k}^{(0)}=\hspace{27mm} \nonumber \\
\frac{k_F}{m_r}\,\left(-\sum_{K} \alpha _{Kq}^{(1)}+\sum_{Q} \alpha _{kQ}^{(1)}-n\us\alpha_{kq}^{(1)}\right)
\label{eq14}
\end{eqnarray} 
Now we can sum this last equation Eq.(\ref{eq14}) over $q$, which gives $n\us\left(E_b+k^2/(2m_r)+(1/r-1)E_F/3\right)\alpha_k^{(0)}=E_b$, taking into account Eq.(\ref{eq13}). Using the relation $\sum_{k} \alpha _{k}^{(0)}=-1$, we get a closed equation for $E_b$ which can be easily solved numerically. For equal masses $r=1$, it gives $E=-E_b=1.389\,E_F$ while from Eq.(\ref{muex}) the exact result is $E=-E_b=E_F$. One can see that working to the second order approximation does not bring any change
to this result, as supported by numerical results. However we will not give details on this since this matter is not easy and not a major interest.

\section{Unequal masses}
\begin{figure}
\includegraphics[width=\linewidth]{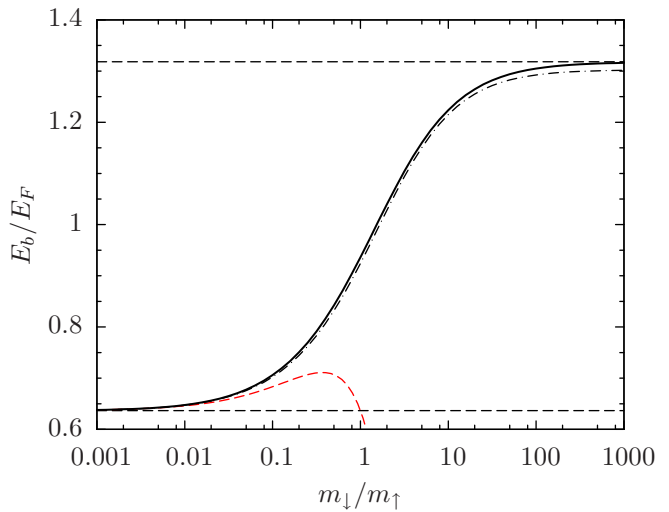}
\caption{\label{fig6} (Color online) Reduced $\down$-atom binding energy $E_b$ as a function of $r=m\ds/m\us$ for $m\us g/k_F=-1$.
Dashed-dotted line: first order approximation. Solid thick line: second order approximation. For $r\to\infty$, $E_b$ goes to $(1+1/\pi)\,E_F\simeq 1.318\,E_F$ (Eq.(\ref{muinf})), while for small $r$, $E_b/E_F\simeq 2/\pi-2r\log r/\pi^2$ (dashed red line).}
\end{figure}

\begin{figure}
\includegraphics[width=\linewidth]{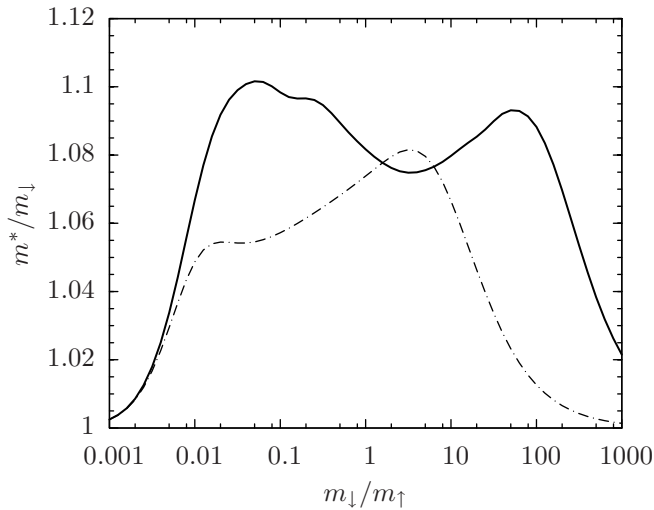}
\caption{\label{fig6b} Relative effective mass $m^\ast/m\ds$ as a function of $r=m\ds/m\us$ for $m\us g/k_F=-1$. Dashed-dotted line: first order approximation. Solid thick line: second order approximation. For $r\to 0$ and $r\to\infty$, $m^\ast$ goes to $m\ds$.}
\end{figure}

In the following, we come back to the general case of unequal masses. We restrict ourselves to
the attractive case and consider as an example the single value $m\us g/k_F=-1$ for the coupling constant.
The results of our calculations as a function of the mass ratio $r=m\ds/m\us$ are presented in Fig.\ref{fig6} and \ref{fig6b}, respectively for the binding energy and the effective mass. For this value of
the coupling constant the binding energy does not display very large variations. The limiting cases
can be understood fairly easily. 

For $r \rightarrow 0$, at fixed coupling constant $g$, the binding energy $E_b$ goes to $-gn\us$. Indeed in this case, the kinetic energy
of the $\down$-spin particle goes to infinity as soon as its momentum is not zero. This implies that $E^{(1)}_{p,kq}$ and $E^{(2)}_{p,kk'qq'}$ go to infinity,
so that $\alpha _{kq}$ and $\alpha _{kk'qq'}$ go to zero in Eq.(\ref{eq1}) and Eq.(\ref{eq2}). 
Clearly this argument could be extended to any order in our successive approximation scheme.
Hence this result is exact. This is quite reasonable physically since this limiting situation can also
be seen as the case of a single particle of mass $m\ds$ moving in the presence of $n\us$ particles
of infinite mass, which means that there is no dynamics associated with them. Hence it is quite natural to find only the mean-field term $-gn\us$ in the binding energy.

The first correction is obtained from Eq.(\ref{eq1}). To dominant order, we find $\alpha_{kq}=-g\alpha_0/\left[E(q-k)+\epsilon_k-\epsilon_q\right]$. Inserting this relation in Eq.(\ref{eq0}) gives $E_b=-gn\us-g^2m\us r\log r/\pi^2$.

As seen from Fig.\ref{fig6b}, the variations of the effective mass are very small, which is the essential
result. It is not clear how the detailed structure would survive in improved approximations. The limiting
cases are again easy to obtain. For the same reason as above, the effective mass in the limit 
$r \rightarrow 0$ is merely equal to the bare mass $m^{\ast}=m\ds$. And the same is obviously
true in the opposite case where $r \rightarrow \infty$, which we consider now with respect to
the binding energy as a function of the coupling constant.

\section{Infinite mass} 

We consider now in more details the other limit where the mass of the $\down$ particle goes to infinity $m\ds\to\infty$. In this case, there is no dynamics associated
with this particle and the problem reduces to the one-body problem of an impurity interacting with a free Fermi sea through a contact potential. It can be solved exactly in a way analogous to the 3D case. The impurity is at the origin and the system is enclosed in the segment $[-R,R]$, with $R\to \infty$ in the thermodynamic limit. Without loss of generality we assume for simplicity that the number of $\up$-spin atoms $N$ is even. The only scattered states correspond to even wave functions $\varphi_k^+(x)$, while odd wave functions $\varphi_k^-(x)$ do not feel the contact potential and behave just as free
particles:
\begin{eqnarray}
\varphi_k^+(x)&=&A_k \cos (k|x|+\delta(k)) \\
\varphi_k^-(x)&=&\sqrt{1/R} \sin (kx)
\end{eqnarray}
where the phase shift is given by $\cot \delta(k)=ka$  and $A_k^2=1/R$ to order $1/R^2$ which is sufficient for our purposes. 

\begin{figure}
\includegraphics[width=\linewidth]{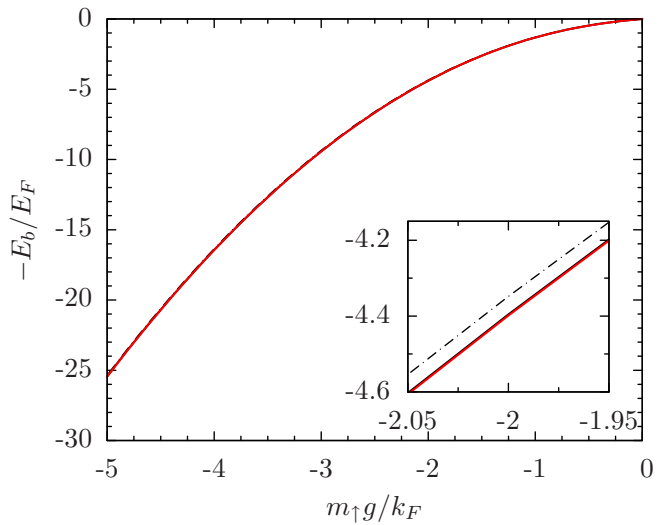}
\caption{\label{fig7} (Color online) Reduced $\down$-atom binding energy $E_b$ as a function of $m\us g/k_F$ for $m\ds=\infty$ and attractive interactions $g<0$. From top to bottom: numerical results for the first (dashed-dotted line) and second (solid thick line) order approximations and exact  solution (Eq.(\ref{muinf})) (solid red line). The inset shows a zoom of the region around $m\us g/k_F=-2$. }
\end{figure}

\begin{figure}
\includegraphics[width=\linewidth]{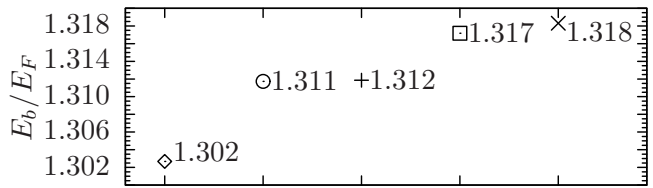}
\caption{\label{fig8} Reduced $\down$-atom binding energy $E_b$ for $m\ds=\infty$ and $m\us g/k_F=-1$. Successive approximations to the exact many-body problem discussed in the text are applied. Diamond: first order approximation. Circle: second order with $qq'=00$. Plus: second order with $qq'=q0$. Square: second order with no $q$ approximation. Cross: exact solution.}
\end{figure}

Since the wave function has to be zero for $x=R$,  the allowed wave vectors $k_p^+$ and $k_p^-$ are given by $k_p^+R+\delta(k_p^+)=(p-1/2)\pi$ and $k_p^-R=p\pi$.
For positive coupling constant $g$, the $N$ $\up$-spin atoms occupy $N$ scattering states ($N/2$ odd states and $N/2$ even states) corresponding to $p=1,\cdots,N/2$. Thus, the Fermi momentum is given by $k_FR=(N/2)\pi$ and the energy of all the atoms is $E=\sum_{p=1}^{N/2} {(k_p^+)}^2/2m\us+\sum_{p=1}^{N/2} {(k_p^-)}^2/2m\us$. For negative coupling constant $g$, our $\delta$-potential also supports a bound state with binding energy $\epsilon _b= 1/2m\us a^2=m\us g^2/2$ and its wave function is $\psi(x)=e^{-|x|/a}/\sqrt{a}$. Starting from the non-interacting situation, we conclude that the $N$ $\up$-spin atoms occupy $N/2-1$ even scattering states corresponding to $p=2,\cdots,N/2$, the single bound state and $N/2$ odd states. Therefore, the energy of all the atoms is $E=-1/2m\us a^2+\sum_{p=2}^{N/2} {(k_p^+)}^2/2m\us+\sum_{p=1}^{N/2} {(k_p^-)}^2/2m\us$. 

We then identify the difference between the non-interacting Fermi sea energy and this energy with the $\down$-atom binding energy.
Since the odd states are not scattered, we find to order $1/N$ :
\begin{eqnarray}
\frac{E_b}{E_F}=y^2\theta(-y)+\frac{8}{\pi N^2}\sum_{p=1}^{N/2}p\, \delta(k_p^+)
\end{eqnarray}
where $y=m\us g/k_F$ and $\theta$ is the usual Heaviside function. To order $1/N$, the phase shift writes $\delta(k_p^+)=-\cot^{-1}(2p/(yN))$. We convert now this sum to an integral and find:
\begin{eqnarray}
\frac{E_b}{E_F}=y^2\theta(-y)-\frac{2}{\pi}\int_0^1 u\cot^{-1}\left(\frac{u}{y}\right)du.
\end{eqnarray}
Finally, we obtain:
\begin{eqnarray}\label{muinf}
\frac{E_b}{E_F}=-\frac{1}{\pi}\left[y-\frac{\pi}{2}y^2+(1+y^2)\arctan y\right]
\end{eqnarray}
This result is valid both for positive and negative $g$, which is not obvious since there is a bound state in the latter situation but not in the former one. This situation is the same as the one found for equal masses \cite{mcg}. Note also that this result is just one half of the exact result for equal masses 
Eq.(\ref{muex}), provided we take in the definition of $y$ the reduced mass, i.e. $y=m_r g/k_F$. 

\begin{figure}
\includegraphics[width=\linewidth]{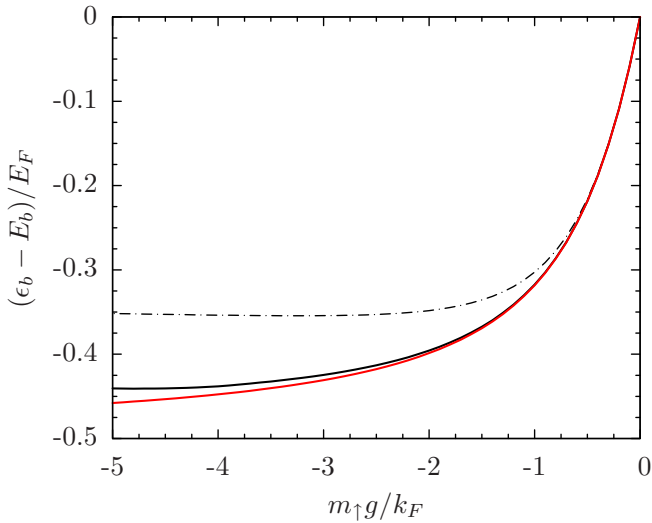}
\caption{\label{fig9} (Color online) Reduced $\down$-atom binding energy $E_b$ as a function of $m\us g/k_F$ for $m\ds=\infty$. Same conventions as in Fig.\ref{fig7}.}
\end{figure}

\begin{figure}
\includegraphics[width=\linewidth]{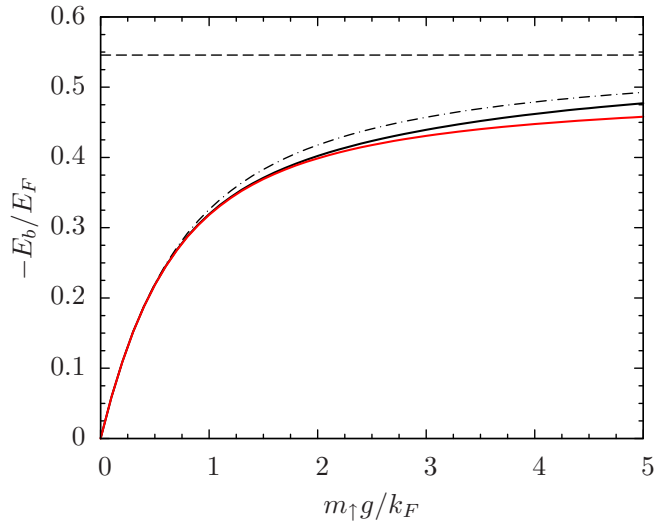}
\caption{\label{fig10} (Color online) Reduced $\down$-atom binding energy $E_b$ as a function of $m\us g/k_F$ for $m\ds=\infty$ and repulsive interactions $g>0$. In the strong repulsive limit $g\to\infty$, first and second order approximations give $E=-E_b=0.5457\,E_F$ while the exact result is $E=-E_b=0.5\,E_F$. Same conventions as in Fig.\ref{fig7} are used.}
\end{figure}

Here again, we can check the convergence of our theoretical scheme \cite{cg}. This is done in Fig.\ref{fig7}, which is very similar to Fig.\ref{fig1}. Similarly the inset shows a zoom around $m\us g/k_F=-2$, where it is clearly seen again that, although the first order result is already quite good, it is markedly improved by the second order one. Specifically for $m\us g/k_F=-1$, the first and second order approximations give respectively $E_b=1.302\,E_F$ and $E_b=1.317\,E_F$ while the exact solution is $E_b/E_F=1+1/\pi \simeq 1.318$. We also show in Fig.\ref{fig8}, together with these values, the results for the intermediate approximations, mentioned above for equal masses.

Considering the large coupling constant limit, we remove again from the binding energy the trivial contribution $\epsilon _b$ of the bound state and display the result in Fig.\ref{fig9}. As it can be seen from Eq.(\ref{muinf}), the exact result is $E_b-\epsilon _b=0.5\,E_F$ in this limit, with a corrective term $-(2/3\pi )E_F/|y|$. Although the relative error is quite small in this regime, as it can be seen from Fig.\ref{fig9}, the absolute error becomes larger, of order of $0.1$, for the first as well as for second order approximation. This is clearly due to the near singularity present in the equations, linked to the existence of the bound state.

The corresponding results are given in Fig.\ref{fig10} for positive coupling constant. Just as for the case of equal masses, the approximations get worst when the coupling constant increases, but the deterioration is markedly less pronounced. For $g\rightarrow \infty$ the exact result is easily found to be $E=-E_b=0.5\,E_F$. The variation of the binding energy when $g$ goes from $+\infty$ to $-\infty$ is therefore $\epsilon _b+E_F$. This is easy to understand.
The occupied plane waves have exactly the same wave vectors in the two limits (the phase shift is $\pi /2 \;({\rm mod}\, \pi )$ in both cases). The only difference is that, for $g \rightarrow -\infty$ we have the bound state with binding energy $\epsilon_b$, and in addition a $\up$ particle has been transferred to this bound state from the Fermi level, which explains the additional contribution $E_F$. It is worth noticing that this variation is half the one found \cite{mcg} in the case of equal masses.

\begin{figure}
\includegraphics[width=\linewidth]{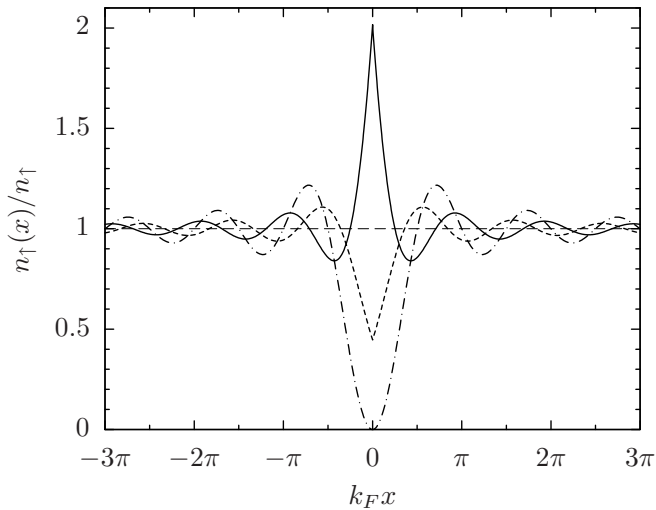}
\caption{\label{fig11} Reduced $\up$-spin local density as a function of $k_Fx$ for various coupling constants $y=m\us g/k_F$. Solid line: $y=-0.5$, thin dashed line: $y=0$, dashed line: $y=0.5$, dashed-dotted line: $y=\pm\infty$ (in the case $y=-\infty$ there is in addition a $\delta$ function spike at $x=0$).}
\end{figure}

Finally it is easy and of interest to calculate the $\up$-spin local density from the above considerations.
It is given by the sum of absolute squares of the $N$ wave functions. Converting this sum to an integral yields:
\begin{eqnarray}\label{nup}
& &\frac{n_\up(x)}{n_\up}=1-\pi y e^{2k_F|x|y}\theta(-y)\\
& &+\int_0^1\left[\cos^2\left(uk_F|x|\!-\!\cot^{-1}(u/y)\right)-\cos^2\left(uk_Fx\right)\right]du.\nonumber
\end{eqnarray}
For $y=0$ one has naturally $n_\up(x)=n_\up$, while for $y\to \pm \infty$ this result gives $n_\up(x)=n_\up\left[1-\sin(2k_Fx)/(2k_Fx)\right]$ which vanishes when $x=0$. Note that this exact result Eq.(\ref{nup}) for infinite $m\ds$ is significantly different from the exact result found for equal masses \cite{mcg}. In this latter case, the $\up$-spin atoms density is always lower than $n_\up$ for positive $g$ and greater than $n_\up$ for negative $g$ \cite{remarq}.
The resulting $n_\up(x)/n_\up$ is displayed in Fig.\ref{fig11} for several values of $y$.

\section{conclusion}

We have considered in this paper the problem of a single particle ("impurity") in the presence of $N$ identical fermions, at zero temperature and in the one-dimensional case.
The only interaction term is between the impurity and the $N$ fermions, which have no interaction between themselves as suited for the effective
interaction of a single species of ultracold fermions.
This situation can easily be realized experimentally in ultracold Fermi gases. This problem can be considered as an extreme case of polarized Fermi gases
and its solution is relevant for the case of strong polarization. We have used an approximate method developed in the three-dimensional case, where
the Hilbert space for the excited states of the $N$ fermions is restricted to have at most two particle-hole pairs. We have compared our approximate
results with exact ones, which are known for special values of the mass ratio. When the mass of the impurity is equal to the mass of the fermions,
we find an excellent agreement for the binding energy as well as the effective mass in the attractive case. For the repulsive case, which has not been
considered in three dimensions, the agreement is less satisfactory, but this easily understood. The other exactly soluble case is the one where the mass
of the impurity is infinite. The problem reduces to the one-body problem of a static impurity interacting with a free Fermi sea, which we solve exactly.
Once again the agreement of our approximate results with the exact ones is excellent. This overall agreement in the one-dimensional case brings a strong validation
for our method applied in three dimensions. Moreover it provides an excellent approximation for the one-dimensional problem in the general case
where the mass of the impurity is not simply related to the fermion mass.

We are grateful to X. Leyronas for stimulating discussions. The ``Laboratoire de
Physique Statistique'' is ``Laboratoire associ\'e au Centre National de la Recherche
Scientifique et aux Universit\'es Paris 6 et Paris 7''.


\begin{references}
\bibitem{remarq}At first this feature seems unphysical since it is apparently in violation of particle conservation.
Actually one can see that, for a finite system with size $2R$, particle conservation is satisfied. However the oscillations around $n\us$, necessary for this conservation, occur for larger and larger distance from the origin when $R$ grows. Hence they are not present at a finite distance in the thermodynamic limit $R \to \infty$.
\end{references}
\end{document}